# Scalability of Atomic-Thin-Body (ATB) Transistors Based on Graphene Nanoribbons

Qin Zhang[1], Yeqing Lu[1], Huili Grace Xing[1], Steven J. Koester[2], and Siyuranga O. Koswatta[3]

[1]Department of Electrical Engineering, University of Notre Dame, Notre Dame, IN 46556, USA

Tel: (574) 631-5498, Fax: (574) 631-4393, email: qzhang1@nd.edu

[2]University of Minnesota, 200 Union St. SE, Minneapolis, Twin Cities, MN 55455, USA

[3]IBM Research Division, T. J. Watson Research Center, Yorktown Heights, NY 10598, USA

## Abstract

A general solution for the electrostatic potential in an atomic-thin-body (ATB) field-effect transistor geometry is presented. The effective electrostatic scaling length, $\lambda_{eff}$, is extracted from the analytical model, which cannot be approximated by the lowest order eigenmode as traditionally done in SOI-MOSFETs. An empirical equation for the scaling length that depends on the geometry parameters is proposed. It is shown that even for a thick $SiO_2$ back oxide $\lambda_{eff}$ can be improved efficiently by thinner top oxide thickness, and to some extent, with high-k dielectrics. The model is then applied to self-consistent simulation of graphene nanoribbon (GNR) Schottky-barrier field-effect transistors (SB-FETs) at the ballistic limit. In the case of GNR SB-FETs, for large $\lambda_{eff}$, the scaling is limited by the conventional electrostatic short channel effects (SCEs). On the other hand, for small $\lambda_{eff}$, the scaling is limited by direct source-to-drain tunneling. A subthreshold swing below 100mV/dec is still possible with a sub-10nm gate length in GNR SB-FETs.

## I. INTRODUCTION

Graphene has been attracting considerable attention as an alternative channel material to silicon in CMOS technology [1]-[6] because of its remarkable electrical properties [1] as well as due to the fact that an energy band gap can be induced under certain conditions in this normally gapless material [2]-[4]. Therefore, it may be possible to realize atomic-thin-body (ATB) transistors based on graphene. To date, GNR based transistor operation has been numerically simulated using the nonequilibrium Green's function formalism combined with 3D electrostatics [5]-[6]. A generalized scaling theory for the ATB field-effect transistor (ATB-FET) geometry however, has not been presented, which is different from the well established ultrathin-body (UTB) structures which have comparatively large body thicknesses with a well-defined semiconductor dielectric medium [7]-[9]. In this letter, a new generalized analytical solution is derived to calculate the electrostatic potential in the ATB-FET geometry which is applicable to transistors based on single-layer and few-layer graphene as well as conventional semiconductor materials with body thicknesses below ~ 1nm [10]. On the other hand, carrier transport properties will depend on the individual bandstructure of the specific material. Here, we simulate GNR based Schottky-barrier FETs (SB-FETs) to explore their scalability dependence on geometrical parameters.

## II. APPROACH

The structure of a double-gate (DG) ATB transistor is illustrated in Fig. 1(a). The atomic thin channel material with a length $L_g$ and a negligible thickness (assumed to be zero in the model) is placed in between the top and back oxides. The top oxide has a thickness of $t_{tox}$ and dielectric constant of $\varepsilon_{tox}$, and the back oxide is $SiO_2$ ($\varepsilon_{box} = 3.9\varepsilon_0$) with a thickness of $t_{box}$. The potential boundary conditions are as shown in Fig. 1(a).

It should be noted that due to the negligible body thickness of ATB-FETs the parabolic approximation for the potential across the body cannot be used here as was done in original scaling studies [11]-[12]. Therefore, we use a procedure similar to that in [7]-[9], whereby the electrostatic potential of the ATB-FET can be written as $\psi(x,y) = v(x) + ul(x,y) + ur(x,y)$, where $v(x)$ is the long-channel solution that accounts for the top and bottom boundary conditions, as well as the self-consistent charge induced in the channel (therefore, the solution is valid in both subthreshold as well as above-threshold operation). $ul(x, y)$ and $ur(x, y)$ satisfy the source and drain boundary conditions, respectively, and can be expanded as:

$$ul(x,y) = \begin{cases} \sum_{n=1}^{\infty} A_n^l \dfrac{\sin[k_n(x+t_{tox})]}{\sin(k_n t_{tox})} \dfrac{\sinh[k_n(L_g - y)]}{\sinh(k_n L_g)} (-t_{tox} \le x \le 0) \\ \sum_{n=1}^{\infty} - A_n^l \dfrac{\sin[k_n(x-t_{box})]}{\sin(k_n t_{box})} \dfrac{\sinh[k_n(L_g - y)]}{\sinh(k_n L_g)} (0 \le x \le t_{box}) \end{cases} \tag{1}$$

$$ur(x,y) = \begin{cases} \sum_{n=1}^{\infty} A_n^r \dfrac{\sin[k_n(x+t_{tox})]}{\sin(k_n t_{tox})} \dfrac{\sinh(k_n y)}{\sinh(k_n L_g)} (-t_{tox} \le x \le 0) \\ \sum_{n=1}^{\infty} - A_n^r \dfrac{\sin[k_n(x-t_{box})]}{\sin(k_n t_{box})} \dfrac{\sinh(k_n y)}{\sinh(k_n L_g)} (0 \le x \le t_{box}) \end{cases} \tag{2}$$

where the eigenvalues $\lambda_n = 1/k_n$ are determined by the implicit equation, $\varepsilon_{box}\tan(k_n t_{tox}) + \varepsilon_{tox}\tan(k_n t_{box}) = 0$. The coefficients $A_n^l$ and $A_n^r$ are calculated by integrating $ul(x, 0)*g_n(x)$ and $ur(x, L_g)*g_n(x)$ from $-t_{tox}$ to $t_{box}$, respectively, where $g_n(x)$ is constructed to be a set of corresponding conjugate functions to each term in $ul(x, 0)$ that satisfies $\int_{-ttox}^{tbox} ul_n(x,0) g_m(x) dx = \delta_{n,m}$ (note that $g_n(x)$ will simultaneously satisfy a similar condition with each term of $ur(x, L_g)$ as well). The calculated eigenvalues $\lambda_n$ and coefficients $A_n^l$, $A_n^r$ are shown in Fig. 1(b). Using the above series solution, an effective scaling length ($\lambda_{eff}$) is extracted by an exponential fitting of the potential profile near the source region. Since the eigenmodes are weighted by the expansion coefficients $A_n^l$ and $A_n^r$, which do not necessarily peak at $n = 1$ (Fig. 1(b)), $\lambda_{eff}$ *cannot* be approximated by the first eigenvalue ($\lambda_1$) as has been traditionally done in SOI-MOSFETs [7]-[9]. For the geometry considered in Fig. 1(b) the first eigenvalue gives $\lambda_1 \approx$ 16nm, which shows a large discrepancy to the 2D series solution with $\lambda_{eff} \approx 1.6$nm.

We use the above series solution to simulate GNR SB-FETs, considering a 2nm-wide GNR with an energy band gap of $E_g = 0.69$eV [13], and source/drain SB height of $\Phi_B = 0.1$V. The Schottky barrier and direct source-to-drain (S→D) tunneling are both included on the same footing, whereby the WKB tunneling coefficient is calculated using the imaginary bandstructure of the GNR that treats the electron-hole nature of carriers realistically. Ballistic transport simulations are performed similar to [14] accounting for self-consistent electrostatics, while any edge-induced states in the band gap [15] and phonon scattering [16] are not considered.

## III. RESULTS AND DISCUSSION

Figure 2(a) compares the extracted $\lambda_{eff}$ to the lowest order eigenvalue $\lambda_1$ as a function of $t_{tox}$ for different $t_{box}$ values. It is clearly seen that $\lambda_1$ cannot correctly describe $\lambda_{eff}$. Therefore, an empirical equation $\lambda_{emp} = \alpha t_{tox}\left[1+\frac{\varepsilon_{box}}{\varepsilon_{tox}}\right]\left[1+\left(\frac{\varepsilon_{box}}{\varepsilon_{tox}}\right)\left(\frac{t_{tox}}{t_{box}}\right)\right] \Big/ \left[1+\frac{t_{tox}}{t_{box}}\right]$ is proposed to approximate $\lambda_{eff}$, and this equation shows reasonable consistency for a large range of $t_{tox}$ and $\varepsilon_{tox}$ values at thick $SiO_2$ box limit where $\alpha = 0.6$ is the only fitting parameter (Fig. 2). It is seen in Fig. 2(a) that $\lambda_{eff}$ is greatly improved by thinner $t_{tox}$, and also modulated by $t_{box}$ when it is below a certain thickness (< 50 nm). On the other hand, $\lambda_{eff}$ and $\lambda_{emp}$ do not appreciably depend on $t_{box}$ for thicknesses > 50nm. Figure 2(b) plots $\lambda_{eff}$ as a function of $\varepsilon_{tox}$ for different $t_{tox}$, and shows that $\lambda_{eff}$ is also improved by high-k top oxide, but the improvement saturates at a certain point [5] [7]. More importantly, even for a thick $SiO_2$ back oxide, $\lambda_{eff}$ can be reduced below 2nm by thinning the top oxide thickness.

The transfer characteristics of GNR SB-FETs are shown in Fig. 3(a) for different gate lengths. For the calculated geometry, $I_{on}/I_{off} > 10^3$ is achieved at a supply voltage of 0.3V. Figures 3(b) and 3(c) show the band diagram and the energy-resolved current density in the off-state of the $L_g$ = 10nm and 15 nm devices, respectively. For this geometry which has $\lambda_{eff}$ = 1.45nm, when the channel length is $L_g \geq$ 15nm the subthreshold swing (*SS*) approaches the 60mV/dec limit, but it degrades to 85mV/dec at $L_g$ = 10nm because of direct S→D tunneling (Fig. 3(b)). Here, we extract the *SS* values where $\log_{10}(I_D)$-$V_{GS}$ is nearly linear in Fig. 3(a) at $V_{GS}$ ≈ 0.05V. In order to further elucidate the scaling potential of GNR SB-FETs, Fig. 4(a) plots *SS* vs. $L_g$, and Fig. 4(b) plots *SS* vs. $L_g/(\pi\lambda_{eff})$. Since the 2D potential at the middle of the channel near the thermionic barrier region depends on $exp(-L_g/2.\lambda_{eff})$, previous works [7] [9] [12] have shown that *SS* degrades significantly for $L_g/(\pi\lambda_{eff}) \leq 1.5 \sim 2$ due to excessive thermal leakage.

We, however, identify two distinct limiting conditions for *SS* degradation depending on the device geometrical parameters that are yet practically relevant: 1) for very small $\lambda_{eff}$ values, *SS* degradation is mainly due to direct S→D tunneling, while 2) at larger $\lambda_{eff}$, *SS* degradation is due to conventional electrostatic short-channel-effects (SCEs) [7] [9] [12]. For thin $t_{tox}$ (1nm and 2.15nm) in Fig. 4(a) with small $\lambda_{eff}$, *SS* degradation for $L_g$ below ~ 15nm is due to direct S→D tunneling [5], and not because of electrostatic SCEs. An $SS \leq$ 100mV/dec is still possible for sub-10nm $L_g$ with small $\lambda_{eff}$ values, even though obtaining small off-state currents will be a challenge because of the direct tunneling limit. Nevertheless, narrower GNRs with larger band gaps may alleviate this problem [5]. For thick $t_{tox}$ (5nm and 8nm) in Fig. 4(b), on the other hand, with large $\lambda_{eff}$, *SS* increases to ~100mV/dec when the conventional $L_g/(\pi\lambda_{eff}) \approx 1.5$ condition is met [7] [9] [12]. In other words, in the large $\lambda_{eff}$ limit, scaling is controlled by the electrostatic SCEs. Interestingly, when comparing Fig. 2 with SOI-MOSFET simulations in [9] for a similar device geometry, the GNR channel has smaller $\lambda_{eff}$ than the 5nm thick Si based UTB channel, showing the excellent scaling potential of GNR based ATB-FETs.

## IV. CONCLUSIONS

A generalized 2D analytical electrostatic solution is developed for the ATB-transistor geometry. An empirical equation for the effective scaling length has been proposed which cannot be approximated by the lowest order eigenmode as traditionally done in SOI–MOSFETs. Even for a thick $SiO_2$ back oxide, the scaling length can be efficiently improved by thinner top oxide, and to some extent, by high-k dielectrics. It is seen that the scalability of GNR SB-FETs can be limited either by direct S→D tunneling, or conventional electrostatic SCEs, depending on the geometrical parameters. Compared to conventional UTB-FET geometries, excellent scaling potential of ATB-FETs based on graphene has been demonstrated.

## FIGURE CAPTIONS

Fig. 1: (a) Schematic cross section of a double-gate (DG) ATB transistor. The channel material is buried between the top and the back oxides with zero thickness. (b) Eigenvalues $\lambda_n$ and coefficients $A_n^l$ and $A_n^r$ in the potential expansion (Eq. (1) and (2)) of a DG GNR SB-FET with $t_{box}$ = 50 nm ($SiO_2$), $t_{tox}$ = 2.15 nm (high-k with $\varepsilon_{tox} = 20\varepsilon_0$), and $V_{GS} = V_{DS}$ = 0.3 V.

Fig. 2: (a) The extracted effective scaling length $\lambda_{eff}$, the lowest order eigenvalue $\lambda_1$ and the empirical $\lambda_{emp}$ as a function of $t_{tox}$ for different $t_{box}$. The back oxide is $SiO_2$ and the top oxide is high-k with $\varepsilon_{tox} = 20\varepsilon_0$. (b) $\lambda_{eff}$ and $\lambda_{emp}$ as a function of $\varepsilon_{tox}$ for different $t_{tox}$. The back oxide is 50 nm $SiO_2$.

Fig. 3: (a) Calculated transfer characteristics of DG GNR SB-FETs with different gate lengths at $\lambda_{eff}$ = 1.45 nm. Band diagram and current density per energy in the off-state for a DG GNR SB-FET with (b) $L_g$ = 10 nm and (c) $L_g$ = 15 nm, showing that the off-state leakage is dominated by thermionic current at longer channel length (c), and increased dramatically by direct source-to-drain tunneling at short gate length (b) for this geometry ($\lambda_{eff}$ = 1.45 nm).

Fig. 4: The subthreshold swing (*SS*) vs. (a) $L_g$, and (b) $L_g/\lambda_{eff}$. For thin $t_{tox}$ (small $\lambda_{eff}$), *SS* significantly increases for $L_g$ below 15nm due to direct source-to-drain tunneling, while for thick $t_{tox}$ (large $\lambda_{eff}$), *SS* increases to 100mV/dec when $L_g/(\pi\lambda_{eff}) \approx 1.5$ due to electrostatic short-channel-effects (SCEs), as observed in conventional SOI-MOSFETs [7] [9] [12].

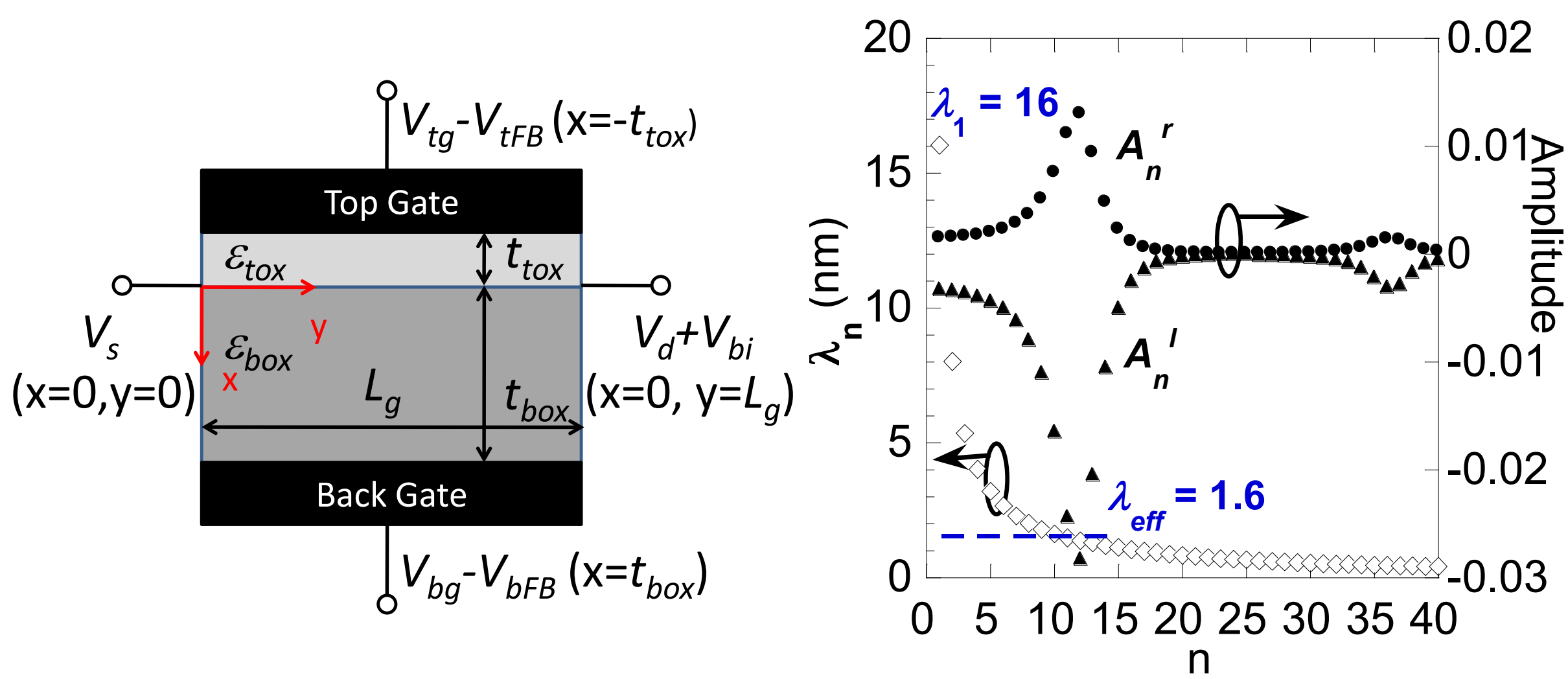

(a) (b)

Fig. 1. Q. Zhang, et al., *IEEE Elect. Dev. Lett.*

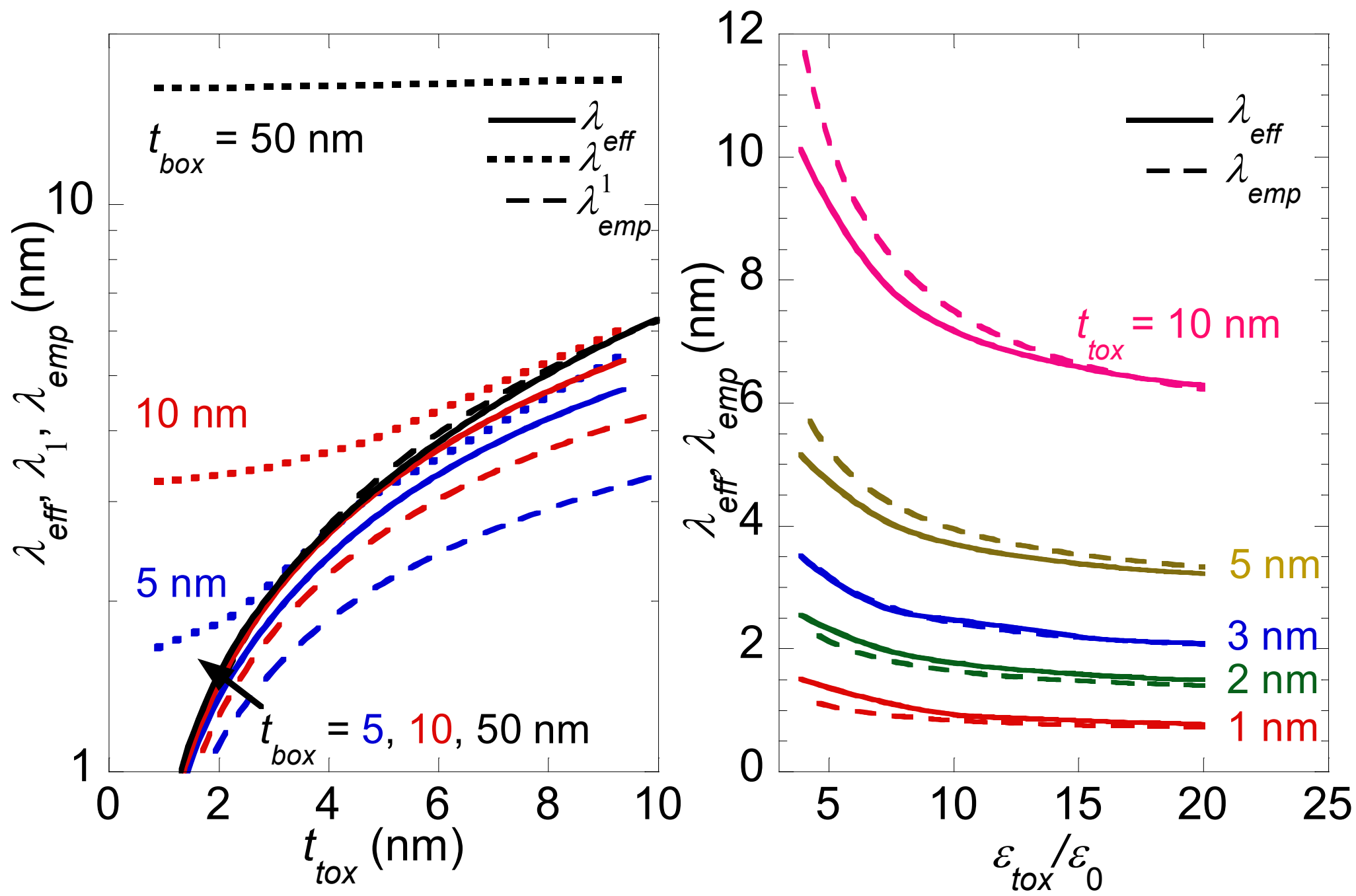

Fig. 2. Q. Zhang, et al., *IEEE Elect. Dev. Lett.*

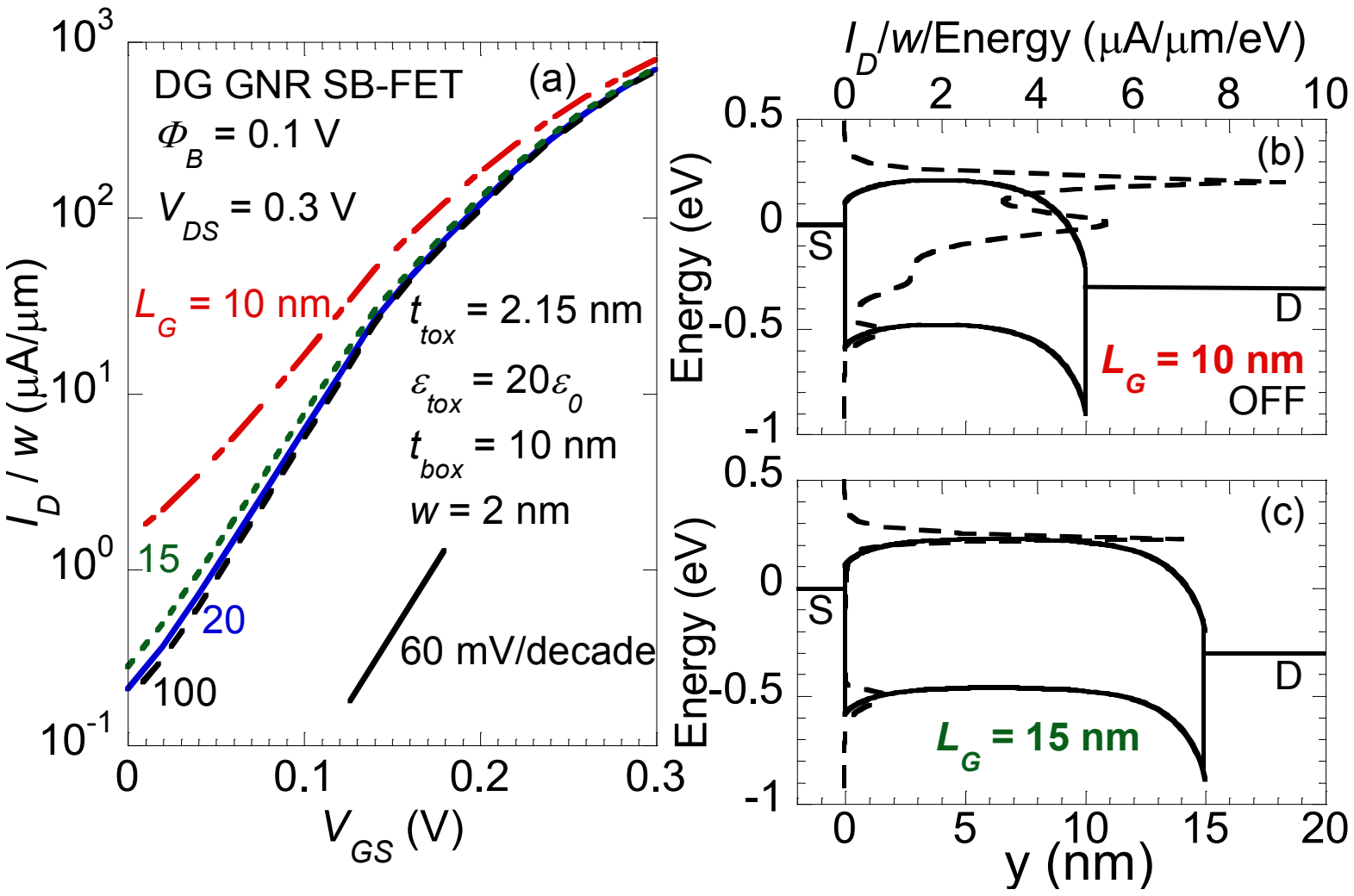

Fig. 3. Q. Zhang, et al., *IEEE Elect. Dev. Lett.*

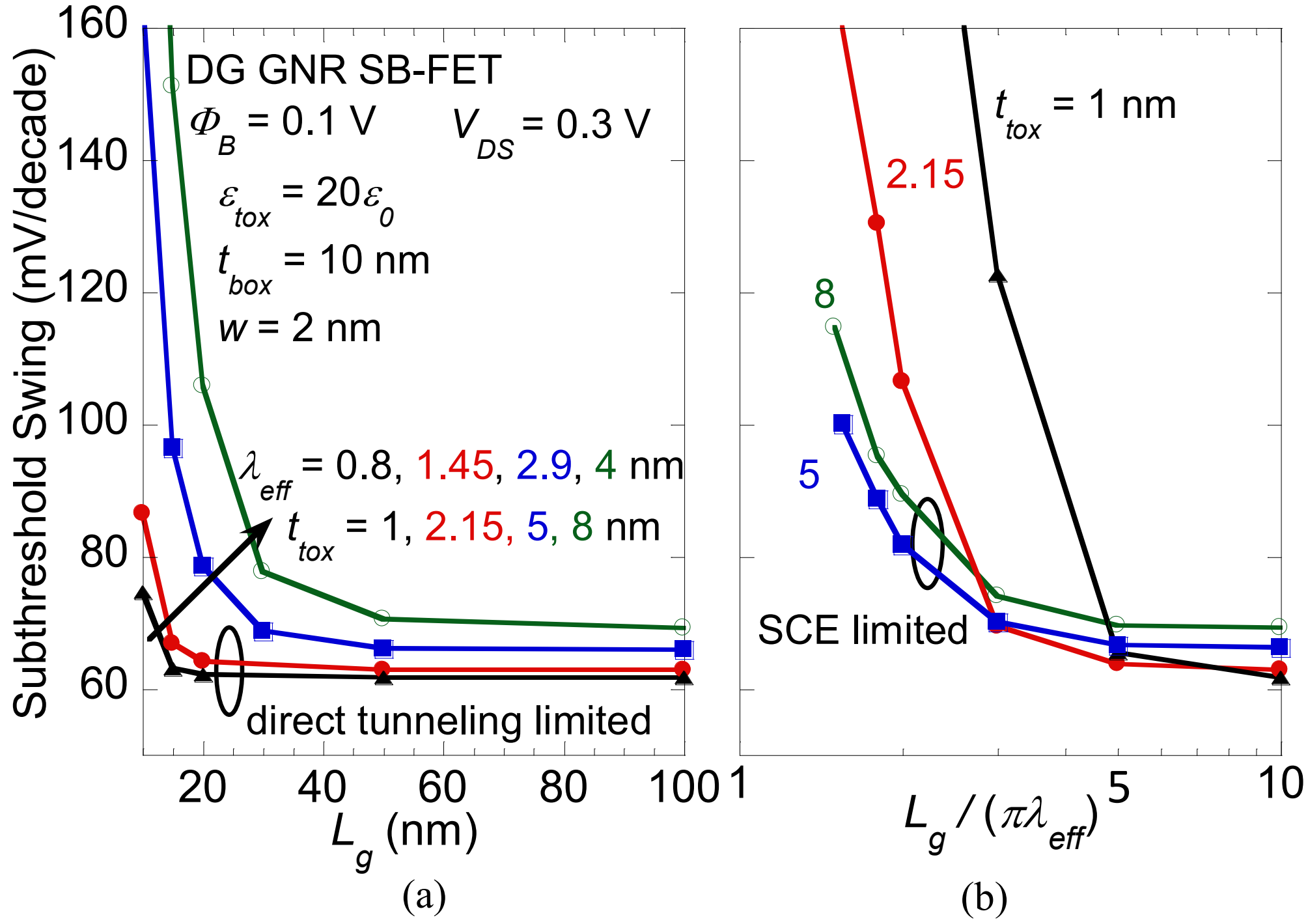

Fig. 4. Q. Zhang, et al., *IEEE Elect. Dev. Lett.*